\documentstyle[prl,aps,twocolumn]{revtex}
\input epsf

\begin{document}
\twocolumn[\hsize\textwidth\columnwidth\hsize\csname@twocolumnfalse%
\endcsname

\title{The magnetic susceptibility of the {\bf $t$-$J$} model
at low hole doping}
\author{D.B. Bailey}
\address{Department of Physics, Stanford University, Stanford, CA 94305}
\author{A. M. Tikofsky\cite{Weiz}}
\address{Institute for Theoretical Physics, University of California,
Santa Barbara, Santa Barbara, CA 93106}
\author{and \\ R. B. Laughlin}
\address{Department of Physics, Stanford University, Stanford, CA 94305\\
        and \\
        Lawrence Livermore National Laboratory,
        P. O. Box 808, Livermore, CA 94550}
\maketitle
\begin{abstract}
We compute the dynamical magnetic susceptibility of the $t$-$J$ model
in its commensurate flux phase at low hole doping.  We compare the
calculations with experiments and exact diagonalization studies. 
\end{abstract}

]
\def\bottomfraction{.9}
\def\textfraction{.1}

It has long been recognized that an essential ingredient in any theory
of high $T_c$ superconductivity is a description of the accompanying
magnetic behavior \cite{magnetic}.  Therefore, a great deal of effort has
been  
invested in studying the magnetic behavior of the $t$-$J$ model,
one of the simplest models thought to capture the physics of
the high temperature superconductors \cite{magnetic2}.
Because the $t$-$J$ model is very strongly correlated, it is not
easily solvable.  Nonetheless, two of the present authors applied a novel
calculational technique to this model and the resulting
optical response function agreed very well with exact diagonalization
studies \cite{tikof,optical}.
In this paper, we extend this calculational formalism in order to
calculate zero temperature magnetic properties
of the $t$-$J$ model.
We find quantitative agreement with exact diagonalization
studies at low dopings and show that our results are
consistent with several experimental properties of the high $T_c$
materials. 

We will study the $t$-$J$ Hamiltonian 
\begin{equation}
{\cal H}_{t-J}  =  -t \sum_{\stackrel{<i,j>}{\sigma}} \, c^{\dag}_{i\sigma}
c_{j\sigma} 
+ \frac{J}{2} \sum_{<i,j>} {\bf S}_i \cdot {\bf S}_j ,
\end{equation}
where $<i,j>$ denotes the Hermitian sum over nearest neighbor pairs, and
no lattice site may be doubly occupied.
In the gauge theory of the $t$-$J$ model, the spin and charge
degrees of freedom are treated as separate while the Gutzwiller constraint
of having no doubly occupied sites is enforced by
a gauge field degree of freedom.   
The corresponding Lagrangian is
\begin{eqnarray}
{\cal L} & = & \sum_j \left\{ \textstyle{\sum_\sigma} f^{\dag}_{j\sigma}
(i\hbar
\frac{\partial}{\partial t} + \phi_j) f_{j\sigma} + b^{\dag}_j(i\hbar
\frac{\partial}{\partial t} + \phi_j) b_j - \phi_j \right\} \nonumber \\
& - & \sum_{<j,k>} \textstyle{ \left\{ -\frac{J}{4}|\chi_{jk}|^2 +
\chi_{jk}\left[ \frac{J}{2} \sum_\sigma f^{\dag}_{j\sigma} f_{k\sigma} + t
b^{\dag}_j b_k \right] \right.} \nonumber \\
& + & \textstyle{ \left. \left[ \frac{t^2}{J}-\frac{J}{8} \right] b^{\dag}_j
b^{\dag}_k b_k b_j \right\}} ,
\end{eqnarray}
\noindent
where $c^{\dag}_{i\sigma} = f^{\dag}_{i\sigma} b_i$, $b^{\dag}_i$ is
a bosonic operator that creates a charge one spinless excitation at site $i$,
and $f^{\dag}_{i\sigma}$ is a neutral operator that creates a spin
$\sigma$ excitation at site $i$.
The time component of the $U(1)$ gauge field at site $j$ is given by
the Lagrange multiplier $\phi_j$ while the phase of the Hubbard-Stratonivich
variable $\chi_{jk}$ is the integral of the spatial component of this
gauge field along the link joining sites $j$ and $k$.

Our calculational procedure requires us to define a saddle point of this
gauge theory \cite{ioffe}.
We first define the mean-field
Hamiltonians for the Fermi and Bose sectors 
\begin{equation}
{\cal H}_b = -t \chi_{\rm o} \sum_{<j,k>} e^{i a_{jk}} b^{\dag}_j b_k 
\end{equation}
\begin{equation}
{\cal H}_f = - \frac{J\chi_{\rm o}}{2} \sum_{\stackrel{<j,k>}{\sigma}}
 e^{i a_{jk}} f^{\dag}_{j\sigma} f_{k\sigma}
\end{equation}
\noindent
where we have assumed that 
$\chi_{jk}=\chi_{kj}^{\dag}$ has a link-independent 
magnitude $\chi_{\rm o}$
given by
\begin{equation}
\chi_{\rm o}^2 = -\frac{1}{JN}\langle 0\! \mid {\cal H}_f+ 
{\cal H}_b \mid \! 0 \rangle\ . 
\end{equation}
\noindent
It is important to note that the value of
$\chi_{\rm o}$ is twice that predicted by the saddle point procedure.  This 
bandwidth enhancement is a known effect of Gutzwiller
projection\cite{zhang}. 
The commensurate flux saddle point is defined by choosing 
mean-field values for $a_{jk}$, the value of the gauge field along
the link joining sites j and k,
such that each plaquette of the lattice encloses $(1-\delta)/2$ flux
quanta, where $\delta$ is the percentage of empty sites.  We are
able to investigate small fluctuations about
these mean-field values for the flux per plaquette 
by employing perturbation theory.
One of the reasons we choose the commensurate flux saddle point
is that it minimizes the total energy of these mean field 
Hamiltonians in the relevant doping range \cite{cflux}.
For the fermions,  ${\cal H}_f$ is a
Hofstadter Hamiltonian\cite{hofst} with $M/N = (1-\delta)/2$ flux quanta
per 
plaquette ($M$, $N$: relatively prime integers).  In order to 
perform calculations in the bose sector, a statistical transmutation is
employed that allows us to treat the bosons in their fermionic
representation \cite{anyons}.
This assumption stabilizes the theory at zero temperature because
the bosons are no longer degenerate and have a
corresponding mean-field Hamiltonian with an energy gap.
In addition, this approximation is consistent with known
variational results\cite{vars}.  

Having defined the underlying formalism, let us now calculate the
spin correlation function $S({\bf q},\omega)$.
We first calculate the mean field spinon
polarization bubble,
\begin{equation}
\chi_{\bf q}^0(\omega) = \frac{1}{N} \sum_n \left[
\frac{|\langle 0 \! \mid \! {\cal S}_{\bf q}^+ \! \mid \! n \rangle
|^2}{\omega\!-\!E_{n0}\!+\!i\eta} - \frac{|\langle 0 \! \mid \!
{\cal S}_{-{\bf q}}^- \! \mid \! n \rangle
|^2}{\omega\!+\! E_{n0}\!+\!i\eta} \right],  
\end{equation}
\noindent
where ${\cal S}_{\bf q}^+ = \sum_{\bf k} f^{\dag}_{{\bf k}+{\bf q},\uparrow}
f_{{\bf k}\downarrow}$, ${\cal S}_{\bf q}^- = \sum_{\bf k} f^{\dag}_{{\bf
k}+{\bf q},\downarrow} 
f_{{\bf k}\uparrow}$, $E_{n0}=E_n-E_0$, and the sum is carried out over
all excited states $\mid \!n \rangle$ of ${\cal H}_f$.

\begin{figure}
\begin{center}
\leavevmode
\epsfbox{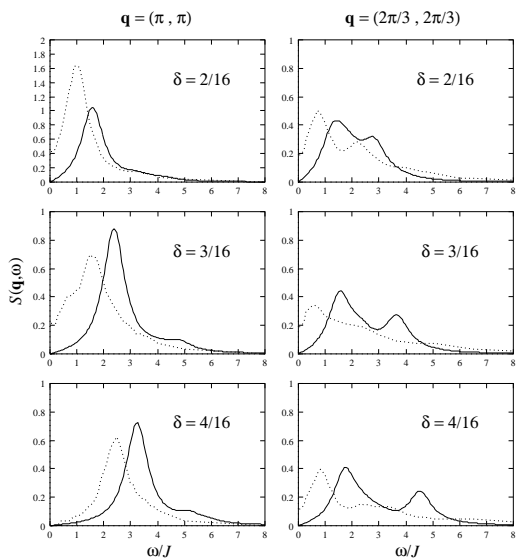}
\end{center}
\caption{
Comparison of $S({\bf q},\omega)$ calculated from Eq. (8) (solid
lines) with the exact diagonalization results of ref. \protect\cite{ederm}
(dashed
lines) at two different momenta for two, three and four holes on a 16-site
cluster for $J/t=0.4$.  Both calculations use $\eta=J/2$.}
\label{fig1}
\end{figure}

Next we perform a vertex correction in order to account for the
gauge field fluctuations that impose the Gutzwiller constraint.
We approximate the gauge field interaction as an instantaneous 
onsite repulsive potential of strength $U$.
We require that $U=1.78 \, J\chi_{\rm o}$, the value found
in variational studies of the Heisenberg antiferromagnet\cite{hsubob}.
This value of $U$, {\it without Gutzwiller projection}, generates the
exact amount of antiferromagnetic order 
which minimizes the energy of the {\it projected} flux ground state.  
We assume that the repulsive core of the gauge field interaction between
spinons is relatively unaffected at low doping.  The reason is that the
core is short-ranged and therefore immune to ``screening'' effects.   
{\it Therefore, $U$ is not an adjustable parameter in this calculation.}
The vertex correction is easily performed,
and the resulting expression for the corrected susceptibility is
\begin{equation}
\label{equ}
\chi_{\bf q}(\omega) = (1-\delta)^2 \frac{\chi^0_{\bf q}(\omega)}{1 + U
\chi^0_{\bf q}(\omega)}, 
\end{equation} 

\noindent
where the doping-dependent prefactor is implicit in the relation
\begin{equation}
{\bf S}_j = \frac{1}{2}f^{\dag}_{j\alpha}[\bbox{\sigma}]_{\alpha\beta}
f_{j\beta} (b_j b^{\dag}_j) \simeq (1-\delta) \frac{1}{2} f^{\dag}_{j\alpha}
[\bbox{\sigma}]_{\alpha\beta} f_{j\beta}\ .
\end{equation}

\begin{figure}
\begin{center}
\leavevmode
\epsfbox{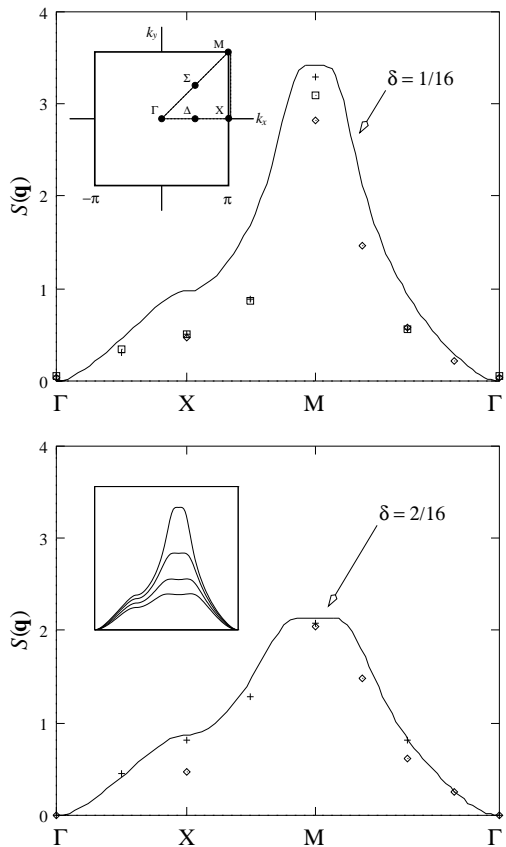}
\end{center}
\caption{
{\it Top}:  the spin structure factor 
$S({\bf q})$ for $\delta = 3/47$ (solid line) compared with the one-hole
exact diagonalization results of ref. \protect\cite{good} ($\diamond$,
$J/t=0.4$) 
and ref. \protect\cite{eder} ($\Box$, $J/t=0.25$; $+$, $J/t=0.5$).  {\it
Inset}: 
Brillouin zone, with special points labeled.  {\it Bottom}:  $S({\bf q})$
for $\delta=5/41$ (solid line) compared with the two-hole results of
ref. \protect\cite{good} ($\diamond$, $J/t=0.4$)  and
ref. \protect\cite{moreo} ($+$,
$J/t=0.4$).  {\it Inset}:  $S({\bf q})$ calculated for (top to bottom)
$\delta \approx 1/16$, 1/8, 3/16, and 1/4.  Note that our $S({\bf q})$ is
independent of $t/J$.} 
\label{fig2}
\end{figure}

Let us now compare our calculation of $S({\bf q},\omega) = -{\rm Im} \,
\chi_{\bf q}(\omega)$ with the exact
diagonalization results of ref. \cite{ederm}.  Fig. \ref{fig1} shows
$S({\bf q},\omega)$ at ${\bf q}={\bf Q}$ (left columns) and ${\bf
q}=\textstyle{\frac{2}{3}}{\bf Q}$ (right) for dopings of 2/16, 3/16 and
4/16 (top, middle and bottom respectively).  The exact diagonalization
values are shown as dotted lines.  The solid lines were obtained from
Eq. (\ref{equ}).  We wish to point out the following basic similarities:

\begin{itemize}
\item There is a single coherent peak with a characteristic energy scale
of $J$ at ${\bf q}={\bf Q}$.  The energy of this pole increases
substantially with doping. 
\item Elsewhere in the zone, (e.g. ${\bf q} = \textstyle{\frac{2}{3}}{\bf
Q}$) the spectral weight is not especially doping-dependent, nor is it
very coherently distributed.
\end{itemize}
\vspace{-0.1in}
\begin{figure}
\begin{center}
\leavevmode
\epsfbox{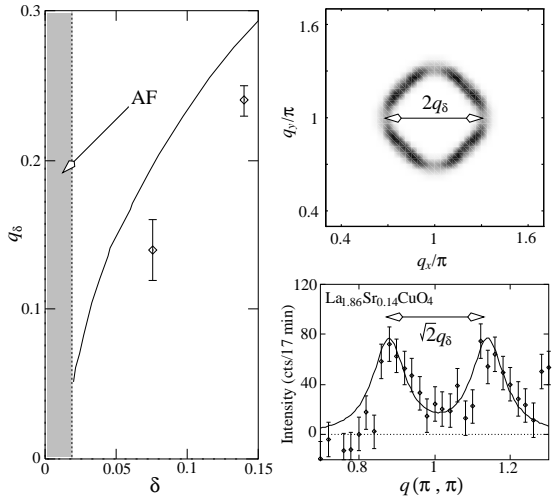}
\end{center}
\caption{
{\it Left}: The discommensuration $q_\delta$ in the lowest energy
spin excitation (solid line) compared with the experimental data of
ref. \protect\cite{cheong}. Due to an antiferromagnetic instability below
$\delta=0.02$ 
(shaded), we have not calculated $q_\delta$ in this regime.  {\it Top
right}: contour plot of $S({\bf q},\omega_0)$ calculated for
$\delta=15/107 \simeq 0.14$ ($J/t$=0.4, $\omega_0=1.14\,J$,
$\eta=0.02\,J$).  Darker shading indicates a larger value
of $S({\bf q},\omega_0)$.  {\it Bottom right}: 
the La$_{1.86}$Sr$_{0.14}$CuO$_4$ neutron scattering data of
ref. \protect\cite{cheong}.  We have 
added the solid line as a guide to the eye.}
\label{fig3}
\end{figure}
\vspace{-0.2in}
\begin{figure}
\begin{center}
\leavevmode
\epsfbox{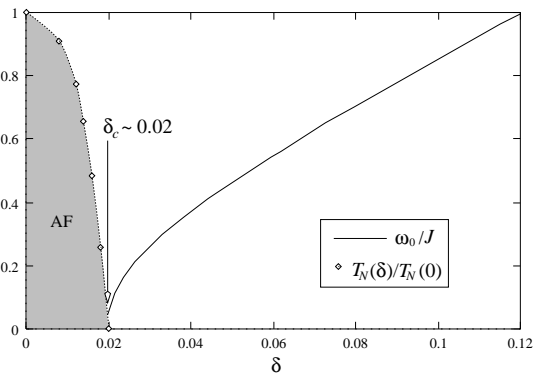}
\end{center}
\caption{
{\it Solid line}: The spin gap frequency $\omega_0/J$ as a
function of doping, for $J/t=0.4$.  
{\it Diamonds} ($\diamond$):  The N\'eel temperature
$T_N$ of the antiferromagnetic phase for
La$_{2-\delta}$Sr$_\delta$CuO$_4$, taken from the experimental data of
ref. \protect\cite{borsa}, for dopings $\delta \leq 0.02$.  The vanishing
spin gap 
frequency as $\delta \rightarrow \delta_c^+$ indicates the instability to
the antiferromagnetic state (shaded) characterized by $T_N$.}
\label{fig4}
\end{figure}

The unphysical sharp structure in both calculations has been smoothed out
considerably through the use of a large value of $\eta$ (= $J/2$).  With
the value of $U$ held fixed, there are no adjustable parameters in our
equation for $S({\bf q},\omega)$.  Furthermore, we have not scaled the $y$
axes.  

We shall now calculate $S({\bf q})=\int_0^\infty d\omega\, S({\bf
q},\omega)$ to see how the integrated spectral weight is distributed over
the Brillouin zone.  This gives an idea of the extent of magnetic
correlation at momentum ${\bf q}$.  Fig. \ref{fig2} compares $S({\bf q})$
with the 
exact diagonalization results of refs. \cite{good,eder,moreo}.  As
expected, the correlation is greatest near ${\bf Q}$ and is suppressed
elsewhere in the zone.  While the exact diagonalization results are weakly
dependent on $J/t$ at ${\bf Q}$, our calculation is independent of J/t.
The doping dependence and
absolute magnitude of $S({\bf q})$ are both reproduced quite well.
Again, we emphasize that there are no adjustable parameters.  

Let us now address the issue of discommensuration in $S({\bf q},\omega)$.
The plateau in
$S({\bf q})$ near ${\bf Q}$ widens upon doping, an effect clearly
visible in the inset in the lower half of Fig. \ref{fig2}.  The widening
plateau is 
an indication of discommensuration of the antiferromagnetic correlations
away from the ordering vector ${\bf Q}$.  The extent of this is usually
characterized by a scalar quantity $q_\delta$, where

\begin{equation}
q_\delta = \frac{1}{\pi} | {\bf q}_d-{\bf Q} |,
\end{equation}

\noindent
and ${\bf q}_d$ is a momentum at which the greatest magnetic scattering
occurs.  Neutron scattering experiments performed on 
La$_{2-\delta}$Sr$_\delta$CuO$_4$ have given values of $q_\delta$ at
$\delta=0.075$ and 0.14 (c.f. Fig. \ref{fig3}, bottom right
panel)\cite{cheong}. 
These values are shown in the left panel of Fig. \ref{fig3}, along with
our 
calculation of $q_\delta(\delta)$.  The upper right panel shows how we
estimate $q_\delta$ from $S({\bf q},\omega)$.  This particular example
shows a doping of $15/107 \simeq 0.14$.  The lowest energy poles in
$S({\bf q},\omega)$ occur at an energy $\omega_0=1.14 \, J$, and are
distributed around ${\bf Q}$ in a ring-like manner (dark shading).  
At higher dopings,
the ring becomes increasingly diamond-shaped, with more spectral weight
along the faces than at the corners.  To be consistent with the
experiments, we take ${\bf q}_d$ in the $(\pi,0)$ direction and calculate
$q_\delta$ as shown.

While the magnitude of the discommensuration does agree with
the experimentally obtained values from La$_{2-\delta}$Sr$_\delta$CuO$_4$,
the distribution does not.  The experiments clearly show that
discommensurate momenta appear at four points in the $(0,\pm\pi)$ and
$(\pm\pi,0)$ directions, not in a ring around ${\bf Q}$.  
This disagreement between our calculation and the experimental data
does not necessarily call into question the validity of our
calculation.  Exact diagonalization studies of the
$t$-$J$ Hamiltonian are as yet unable to 
determine which type of discommensuration it exhibits.
Resolution of this issue will have to wait until clusters of large enough
size can be studied.  For now, we note that the
discommensuration effect that we calculated is due to the existence
of a finite concentration of empty lattice
sites.  Because this distribution is uncorrelated (i.e. there is no charge
order), all directions look equivalent to spin excitations of momentum
${\bf q}$.  Therefore, a ring of discommensurate momenta forms around the
ordering vector $(\pi,\pi)$.  We expect this to change if the ground state
is modified to include axis-aligned charge ordering, a modification which
would require the addition of another Hartree-Fock parameter.  The
formation of such ``stripes'' is thought by some to be an essential
ingredient of high $T_c$ superconductivity\cite{kivem}, and it has indeed
been observed in experiments\cite{stripes}.  If such
ordering were found to occur along the crystalline axes it would cause
discommensuration primarily along the $(\pm\pi,0)$ and $(0,\pm\pi)$
directions.  

It is well known that the cuprate superconductors become antiferromagnets
in the limit of extreme underdoping.  The doping at which this 
occurs has been estimated in several
calculations\cite{delta}.  Our ground state becomes unstable to
antiferromagnetism at dopings below 2\%, which happens to be exactly the
dopant concentration below which long range antiferromagnetic order
appears in La$_{2-\delta}$Sr$_\delta$CuO$_4$\cite{borsa}.  The instability
is indicated by a vanishing gap to spin excitations as
$\delta\rightarrow\delta_c^+$, where $\delta_c \simeq 0.02$.  Above
$\delta_c$, the lowest lying spin excitation requires a finite energy
$\omega_0$.  Fig. \ref{fig4} illustrates the boundary between the stable
regime ($\delta>\delta_c$, $\omega_0>0$) and the unstable regime
($\delta<\delta_c$, $\omega_0=0$).  We have included the values of the
N\'eel temperature measured in ref. \cite{borsa} for
La$_{2-\delta}$Sr$_\delta$CuO$_4$ at dopings $\delta\leq 0.02$.  

While our calculation of the minimum doping of the spin disordered
phase $\delta_c$ agrees with experimental studies of
La$_{2-\delta}$Sr$_\delta$CuO$_4$, our calculation of the
gap to spin excitations $\omega_0$ in this phase does not. 
Neutron scattering investigations of the high $T_c$ materials indicate
a gap to spin excitations with an energy scale
closer to the superconducting gap than to $J$\cite{gap}.
Both our calculation and the exact diagonalization work thus far greatly
overestimate the value of $\omega_0$.  While 
this may be a finite size effect in
the exact diagonalizations, we attribute it to a overly large mean-field
energy gap that is expected to be reduced by broadening
and damping effects such as those considered in
ref. \cite{jared} with respect to the fractional statistics gas.  

To summarize, we have calculated the zero temperature magnetic properties
of the $t$-$J$ Hamiltonian in its commensurate flux phase.  We find
quantitative agreement with exact diagonalization calculations of the spin
correlation function\cite{ederm} and the spin structure
factor\cite{good,eder,moreo}.  We also find discommensuration
of magnetic order away from $(\pi,\pi)$ comparable in magnitude to that
observed in
La$_{2-\delta}$Sr$_\delta$CuO$_4$\cite{cheong}, but the details remain to
be clarified.  The instability to an antiferromagnetic phase occurs at a
value of $\delta$ close to that found in experiments\cite{borsa} and
other calculations\cite{delta}.  However, the gap to spin excitations in the
commensurate flux phase is approximately
five times larger than indicated by experiment \cite{gap}.
Nevertheless, the $t$-$J$ model seems to be 
well described by a gauge theory of new particles identified with the spin
and charge coordinates of the original electrons.  We therefore present
these calculations as a further test of the gauge theory of spinons and
holons and as a justification for continued efforts to remedy its
shortcomings. 

We are indebted to R. Eder for providing us with some of the exact
diagonalization results, and to M. Greiter and E. Demler for useful
discussions.  This work was supported by the NSF under grants PHY94-07194
(A.M.T) and DMR-9421888 (R.B.L).  D.B.B gratefully acknowledges the
support of an NSF fellowship and an ARCS fellowship.


\vspace{-0.2in}
\begin{references}
\vspace{-0.6in}
\bibitem[*]{Weiz}
Present Address:  Dept. of Condensed Matter Physics,\\
Weizmann Institute of Science, Rehovot 76100, Israel. 

\bibitem{magnetic}
P. W. Anderson, {\it Science} {\bf 235}, 1196 (1987); G. Baskaran, Z. Zou,
and P. W. Anderson, {\it Solid State Comm.} {\bf 63}, 973 (1987);
R. B. Laughlin, {\it Science} 242, 525 (1988).

\bibitem{magnetic2}
see, for example, T. Tanamoto, K. Kuboki, and H. Fukuyama, {\it
J. Phys. Soc. Jpn.} {\bf 60}, 3072 (1991);
A. J. Millis, L. B. Ioffe, and H. Monien, {\it J. Phys. Chem. Solids} {\bf
56}, 1641 (1995).

\bibitem{tikof}
A. M. Tikofsky and R. B. Laughlin, Phys. Rev. B {\bf 50}, 10165 (1994).

\bibitem{optical}
A. M. Tikofsky, R. B. Laughlin and Z. Zou, Phys. Rev. Lett. {\bf 69},
3670 (1992).

\bibitem{ioffe}
L. B. Ioffe and A. I. Larkin, Phys. Rev. B {\bf 39}, 8988 (1989).

\bibitem{cflux}
Y. Hasegawa {\it et al.}, Phys. Rev. Lett. {\bf 63}, 907 (1989);
Y. Hasegawa {\it et al.}, J. Phys. Soc. Jpn. {\bf 59}, 822 (1990).

\bibitem{hofst}
D. Hofstadter, Phys. Rev. B {\bf 14}, 2239 (1976);

\bibitem{vars}
R. B. Laughlin, J. Low. Temp. Phys. {\bf 99}, 443 (1995).

\bibitem{anyons}
A. L. Fetter, C. B. Hanna, and R. B. Laughlin, Int. J. Mod. Phys. B {\bf
5}, 2751 (1991).

\bibitem{zhang}
F. C. Zhang {\it et al.}, Superconduct. Schi. Technol. {\bf 1}, 36
(1988). 

\bibitem{hsubob}
T. Hsu, Phys. Rev. B {\bf 41}, 11379 (1990);
R. B. Laughlin, J. Phys. Chem. Solids {\bf 56}, 1627 (1995).

\bibitem{ederm}
R. Eder, Y. Ohta, and S. Maekawa, Phys. Rev. Lett. {\bf 74}, 5124 (1995).

\bibitem{good}
R. J. Gooding, K. J. E. Vos, and P. W. Leung, Phys. Rev. B {\bf 49}, 4119
(1994). 

\bibitem{eder}
R. Eder, unpublished.

\bibitem{moreo}
A. Moreo {\it et al.}, Phys. Rev. B {\bf 42}, 6283 (1990).

\bibitem{cheong}
S. W. Cheong {\it et al}, Phys. Rev. Lett. {\bf 67}, 1791 (1991);

\bibitem{kivem}
V. J. Emery and S. A. Kivelson, Nature {\bf 374}, 434 (1995).

\bibitem{stripes}
J. Tranquada {\it et. al.}, Nature {375}, 561 (1995).

\bibitem{delta}
J. L. Richard and V. Yu. Yushankha\"\i, Phys. Rev. B {\bf 50}, 12927
(1994); 
F. P. Onufrieva, V. P. Kushnir, and B. P. Toperverg, {\it ibid.} {\bf 50},
12935 (1994); G. Khaliullin and P. Horsch, Phys. Rev. B {\bf 47}, 463
(1993). 

\bibitem{borsa}
F. Borsa {\it et al.}, Phys. Rev. B {\bf 52}, 7334 (1995). 

\bibitem{gap}
J. Rossat-Mignod {\it et al.}, Physica B {\bf 192}, 109 (1993);
P. Bourges {\it et al.}, Phys. Rev. B {\bf 43}, 8690 (1991);
J. Tranquada {\it et al.}, Phys. Rev. B {\bf 46}, 5561 (1992).

\bibitem{jared}
J. L. Levy and R. B. Laughlin, Phys. Rev. B {\bf 50}, 7107 (1994).

\end{references}
\end{document}